\documentclass[a4paper,fleqn]{arxiv}

\usepackage{graphicx}
\usepackage{epstopdf, epsfig}
\usepackage{booktabs}
\usepackage{multirow}
\usepackage{array}
\usepackage{calc}
\usepackage{microtype}
\usepackage{lscape}
\usepackage{amsmath,esint,upgreek}
\usepackage{amsmath,amsfonts}

\usepackage[normalem]{ulem}
\usepackage{color}
\usepackage{mylines}
\usepackage{myplots}
\newcommand{\sy}[2]{\mbox{(\kern-.25em\SymbolRGB[solid]{#1}{.8pt}{#2}{5pt}\kern-.25em)}}
\newcommand{\lsy}[3]{\mbox{(\kern-.1em\lineSymbolRGB{#1}{#2}{2pt}{#3}{4pt}\kern-.45em)}}
\newcommand{\lcap}[2]{~\,{\kern-1em\protect\mylcap{#1}{#2}}}

\definecolor{blue}{rgb}{0,0,1}
\definecolor{red}{rgb}{1,0,0}
\definecolor{black}{rgb}{0,0,0}
\definecolor{grey}{rgb}{0.775,0.775,0.775}
\definecolor{white}{rgb}{1,1,1}
\definecolor{OrangeM}{rgb}{0.8500, 0.3250, 0.0980}
\definecolor{BlueM}{rgb}{0, 0.4470, 0.7410}
\definecolor{YellowM}{rgb}{0.9290, 0.6940, 0.1250}
\definecolor{PurpleM}{rgb}{0.4940, 0.1840, 0.5560}
\definecolor{GreenM}{rgb}{0.4660, 0.6740, 0.1880}
\definecolor{LBlueM}{rgb}{0.3010, 0.7450, 0.9330}
\definecolor{Red1}{rgb}{0.8, 0.3, 0.09}
\definecolor{Blue1}{rgb}{0.258823529411765,	0.572549019607843,	0.776470588235294}
\definecolor{Yellow1}{rgb}{0.941176470588235	0.784313725490196	0.0588235294117647}
\definecolor{Green1}{rgb}{0.254901960784314,	0.670588235294118,	0.364705882352941}
\definecolor{redPIV}{rgb}{0.984,0.416,0.290}
\definecolor{bluePIV}{rgb}{0.420,0.682,0.839}
\definecolor{green_laser}{RGB}{0,208,0}
\definecolor{f225_DC75}{RGB}{25,102,174}
\definecolor{f225_DC50}{RGB}{23,127,61}
\definecolor{f225_DC25}{RGB}{202,66,0}
\definecolor{f150_DC75}{RGB}{81,156,204}
\definecolor{f150_DC50}{RGB}{87,181,102}
\definecolor{f150_DC25}{RGB}{245,120,35}

\definecolor{rodri}{RGB}{20, 20, 255}

\usepackage[numbers]{natbib}

\def\tsc#1{\csdef{#1}{\textsc{\lowercase{#1}}\xspace}}
\tsc{WGM}
\tsc{QE}

\begin{document}
\let\WriteBookmarks\relax
\def\floatpagepagefraction{1}
\def\textpagefraction{.001}

\shorttitle{Heat transfer enhancement in TBL with a pulsed  slot jet in CF}    

\shortauthors{R. Castellanos et al.}  

\title [mode = title]{Heat transfer enhancement in turbulent boundary layers with a pulsed slot jet in crossflow}  


\author[1,2]{Rodrigo Castellanos}[
    orcid=0000-0002-7789-5725]
\cormark[1] 
\ead{rcastell@ing.uc3m.es}
\credit{Methodology, Software, Validation, Formal analysis, Investigation, Data Curation, Writing - Original Draft, Writing - Review \& Editing, Visualization, Project administration}
\affiliation[1]{organisation={Aerospace Engineering Research Group, Universidad Carlos III de Madrid},
            city={Legan\'es},
            postcode={28911}, 
            state={Madrid},
            country={Spain}}
\affiliation[2]{organisation={Theoretical and Computational Aerodynamics Branch, Flight Physics Department, Spanish National Institute for Aerospace Technology (INTA)},
            city={Torrej\'on de Ardoz},
            postcode={28850}, 
            state={Madrid},
            country={Spain}}
\author[1,3]{Gianfranco Salih}[
    orcid=0000-0002-3654-5542]
\ead{gianfranco.salih@mail.polimi.it}
\credit{Software, Validation, Investigation, Writing - Original Draft, Writing - Review \& Editing, Visualization}
\affiliation[3]{organisation={Politecnico di Milano},
            city={Milano},
            postcode={20156}, 
            country={Italy}}
\author[1]{Marco Raiola}[
    orcid=0000-0003-2744-6347]
\ead{mraiola@ing.uc3m.es}
\credit{Methodology, Validation, Investigation, Writing - Review \& Editing, Supervision}
\author[1]{Andrea Ianiro}[
    orcid=0000-0001-7342-4814]
\ead{aianiro@ing.uc3m.es}
\credit{Conceptualization, Methodology, Resources, Writing - Review \& Editing, Supervision, Project administration, Funding acquisition}
\author[1]{Stefano Discetti}[
    orcid=0000-0001-9025-1505]
\ead{sdiscett@ing.uc3m.es}
\credit{Conceptualization, Methodology, Validation, Resources, Writing - Review \& Editing, Supervision, Project administration, Funding acquisition}
\cortext[1]{Corresponding author}
\maketitle

\noindent \textsc{Abstract}\\
\\
The convective heat transfer enhancement in a turbulent boundary layer (TBL) employing a pulsed, slot jet in crossflow is investigated experimentally. A parametric study on actuation frequencies and duty cycles is performed. The actuator is a flush-mounted slot jet that injects fluid into a well-behaved zero-pressure-gradient TBL over a flat plate. A heated-thin-foil sensor measures the time-averaged convective heat transfer coefficient downstream of the actuator location and the flow field is characterised by means of Particle Image Velocimetry. The results show that both the jet penetration in the streamwise direction and the overall Nusselt number increase with increasing duty cycle. The frequency at which the Nusselt number is maximised is independent of the duty cycle. The flow topology is considerably altered by the jet pulsation. A wall-attached jet rises from the slot accompanied by a pair of counter-rotating vortices that promote flow entrainment and mixing. Eventually, a simplified model is proposed which decouples the effect of pulsation frequency and duty cycle in the overall heat transfer enhancement, with a good agreement with experimental data. The cost of actuation is also quantified in terms of the amount of injected fluid during the actuation, leading to conclude that the lowest duty cycle is the most efficient for heat transfer enhancement.

\vspace{1cm}
\noindent \textsc{Keywords}\\
\\
Flow control \\ Pulsed flow \\ Crossflow jet \\ Boundary layers \\ Convective heat transfer enhancement

%
\newpage
\section{Introduction}\label{s:intro}
The control of turbulent flows is one of the key research areas of engineering due to its utmost importance in the transport of mass, momentum and heat. Applications such as boundary layer separation control, skin friction reduction or convective heat transfer enhancement motivate the development of methods for turbulent flow control. The majority of available studies focus on the reduction of momentum fluxes in turbulent flows, targeting skin-friction drag; however, convective heat transfer enhancement still raises several challenges, building a technological barrier in fields such as high-power electronics or turbomachinery \citep[see e.g.][]{moore2014emerging,han2012gas}. 

A common solution for heat transfer enhancement in wall-bounded flows consists of producing embedded streamwise vortices in the boundary layer \citep{jacobi1995} by the use of passive elements such as vortex generators \citep{Ke2019VG} or wall-mounted obstacles \citep{nakamura2001cube, mallor2018cubes}.
Passive techniques rely on the interaction between the flow and certain surface geometries. In most cases, the target is the formation of near-wall streamwise vortices which improve the momentum transfer within the boundary layer. These methods are simple, do not require power to operate and need little maintenance. However, passive elements are designed and optimised for a specific working condition, lacking adaptability and introducing intrinsic losses (e.g. the drag and pressure losses induced by obstacles).

On the other hand, active techniques utilise electric or acoustic fields, vibration or flow injection to induce changes in the flow. Such strategies can adapt to the changing conditions, although coming with the cost of spending energy in the actuation. Active methods are not yet widely applied in industrial environments, but there are significant research lines focused on their development due to their theoretical advantage over passive methods.
In the field of heat transfer enhancement, active flow control gathers several strategies such as electro-hydrodynamics forcing \citep[see e.g.][]{ElectroHydroALLEN1995389} to enhance convection by the application of an electric field, plasma actuators to perform a momentum injection that induces turbulent structures modifying the heat transfer capabilities as demonstrated by \citet{castellanos2022plasma} in a turbulent boundary layer or by \citet{roy2008plasma} for film-cooling applications, or impinging jets by exciting the structure and organisation of large scale turbulence \citep{Kataoka1987ij}.
In particular, studies such as the one by \citet{MLADIN19973305} and \citet{LIU19963695} conclude that adding pulsation to an impinging jet can produce an increase or a decrease of the heat transfer capability depending on the actuation frequency. Pulsation at the jet's natural frequency has been observed to promote the formation of small eddies which enhance local heat transfer \citep{LIU19963695}.

Crossflow or transverse jets have also been successfully exploited for control purposes. The jet in crossflow (JICF) is characterised by a counter-rotating vortex pair that forms downstream of the jet \citep{kamotani1972experiments,kelso1996experimental}. This vortex motion determines, for the most part, the velocity and temperature distributions downstream of the jet. According to \citet{fric_roshko_1994}, four types of coherent turbulent structures are present in the near field of the jet: shear-layer vortices which dominate the initial portion of the jet, a system of horseshoe vortices that wrap around the base of the jet, the counter-rotating vortex pair (CVP), and the wake vortices extending from the wall to the jet. The formation and interaction among the coherent structures themselves and with the main flow have been widely studied \citep[see e.g.][]{cortelezzi2001formation, broadwell1984structure} and associated with the enhanced mixing efficiency of crossflow jets \citep{smith1998mixing}.  \citet{broadwell1984structure} asserts that the jet mixing rate becomes independent of the Reynolds number above a critical value, as the flow is dominated by turbulent mixing and the presence of small-scale turbulent motions. It remains however dependent on the velocity ratio between the crossflow and the jet, as it affects the penetration and bend of the injected flow. The velocity ratio also determines the extent of the wake region formed on the lee of the jet, where small velocity gradients and lower mixing are present \citep{andreopoulos1984experimental}. Likewise, \citet{smith1998mixing} describe that the mixing enhancement is limited to the near field region of the flow, dominated by the structural formation of the CVP and not to the far-field region where the CVP is fully developed. 

Regarding heat transfer, \citet{Carlomagno2004heatJICF} experimentally investigated the relations between turbulent statistics of a steady JICF in a laminar boundary layer and the convective heat transfer at the wall, concluding an existing dependency of heat transfer on the in-plane turbulent kinetic energy in the near-wall region. They also pointed out the strong influence of the CVP on heat transfer distribution, which varies with the jet velocity ratio. In this regard, the recent numerical and experimental work by \citet{puzu2019jet} investigated the influence of the counter-rotating vortices induced by a steady JICF in the enhancement of heat transfer. They reported that the maximum mixing and hence heat transfer is achieved some distance downstream of the flow injection, where the CVP merge.
Although most of the studies on jets in crossflow are focused on round jets, there are several contributions (see e.g. \citep{liscinsky1996crossflow,plesniak2005scalar}) describing that the coherent structures and mixing characteristics are not greatly affected by the shape of the injected jet, especially in the far-field region of the flow. Moreover, localised uniform blowing through a spanwise slot in the TBL has also been investigated both experimentally and numerically \citep[see e.g.][]{park1999, Krogstad2000slitjet}.

One of the paths followed to increase the transport of jets in crossflow is introducing a pulsation.
In the work by \citet{vermeulen1992mixing} acoustically-excited jets were injected into a hot crossflow. It was observed that the jet excitation increased the size of the mixing area, penetration, and overall mixing while reducing the required downstream displacement to obtain a certain mixing rate. 
Similarly, \citet{johari1999penetration} studied the mixing properties of fully-modulated (valve-actuated) turbulent jets in crossflow. It was found that both the frequency and duty cycle of the actuation influences the mixing properties, demonstrating that a short injection time provides a higher improvement of mixing and dilution compared to long injection times which resemble more closely steady jets.
\citet{eroglu2001structure} focused on the induced alterations on the vortical structures by the jet pulsation, finding that jets excited by square waves generate distinct vortex rings with up to $70\%$ increase in penetration in comparison to steadily-blowing jets (from now on referred to as ``steady jets'' for brevity). This result motivated \citet{MCLOSKEY2002} to quantify the dynamics of actuation and to develop a control loop for the optimisation of jet penetration and spread through. \citet{Johari2006scaling} investigated the flow structures induced by pulsed JICF, describing the formation of hairpin vortices, vortex rings with or without trailing columns or turbulent puffs depending on the stroke ratio. Recently, \citet{Steinfurth2021pulsedjet} investigated the inclined, slot and pulsed JICF with a rich discussion on the flow topology. They concluded that the flow tends to attach to the wall, forming half-ring vortices that considerably enhance momentum transfer.
Most of the above-mentioned contributions agreed that, by pulsating the jet flow, it is possible to find ideal actions that amplify or optimise a certain characteristic aspect of the flow \citep{shapiro2006optimization}. The objective function for optimisation, however, is not found to be universal and presents great differences due to the dependency on the pulsating conditions, which are found to be apparatus-dependant \citep{MCLOSKEY2002}. The interested reader is referred to the extensive review article by \citet{karagozian2010transverse} covering the jet in crossflow and its control.

For heat transfer purposes, pulsating jets have been widely used in impingement cooling configurations. As an example, \citet{azevedo1994pulsed} studied impingement cooling from a pulsating jet, reporting an overall decreasing effect in comparison with a steady jet. Conversely, \citet{xu2010turbulent} showed that intermittent pulsation jet provides an increase in convective heat transfer over a wide range of frequencies although the study did not focus on the flow mechanisms behind the enhancement. The utilisation of pulsed slot jets in crossflow is, however, barely investigated for heat transfer purposes apart from film cooling applications. \citet{muldoon2009dns} studied the film cooling effectiveness of a pulsating jet injected at a 30$^\circ$ angle with respect to the crossflow at different frequencies and duty cycles. They found an optimal actuation frequency, independent of the duty cycle, at which the enhancement was maximum. This enhancement was then attributed to the enhanced vortex dynamics generated by the pulsation, which then allowed the jet stream to remain attached and spread the flow downwards. Similarly, \citet{coulthard2007effect1, coulthard2007effect2} studied the film cooling enhancement of a streamwise row of 35$^\circ$ inclined pulsating jets. These studies declare that, by pulsating a jet, it is possible to provide equivalent heat transfer rates at reduced film cooling effectiveness, providing however a positive net heat flux effect, while also reducing the mass flow rate into the system.  

Still, the majority of the scientific contributions in the literature focus on mixing enhancement and skin-friction drag reduction. It is worth mentioning the recent work by \citet{zhou2020artificial} exploring genetic programming control to maximise the mixing on a turbulent jet, and the contribution by \citet{cheng2021skin} that combines several slot jets aligned with the freestream to promote skin-friction reduction in a turbulent boundary layer. 
Nonetheless, since convective heat transfer in turbulent flows is dominated by bulk mixing, one could apply the knowledge on mixing enhancement for heat transfer.

To the authors' best knowledge, the performance of pulsed slot jets in crossflow as control devices to enhance convective heat transfer in wall-bounded turbulent flows is still an unexplored field. In this work, we investigate the convective heat transfer features of a pulsed spanwise slot jet in crossflow in an open-loop arrangement, i.e. the actuation laws are prescribed and independent of the state of the flow. A parametric study on two actuation features is carried out: pulsation frequency and duty cycle, i.e. the fraction of time the pulsed jet is ejecting flow between two consecutive pulses. The analysis focuses on describing the effect of both parameters on the wall heat-flux distribution, assessing the extension of the affected area, the absolute performance, and the optimal control strategy within the considered parametric space. A discussion decoupling the contributions of frequency and duty cycle to the heat transfer enhancement is provided based on an \textit{ad hoc} model.

The present article is structured as follows. Section \ref{s:Methodology} defines the experimental apparatus and the measurement techniques employed for the study. Heat-transfer results are collected and described in Section \ref{s:Results} while the discussion of the flow field is provided in Section~\ref{s:pivresults}. Ultimately, the conclusions of the study are drawn in Section \ref{s:Conclusions}.

\section{Methodology and experimental setup \label{s:Methodology}}

\begin{figure*}[t]
    \centering
    \includegraphics[width = 0.99\linewidth]{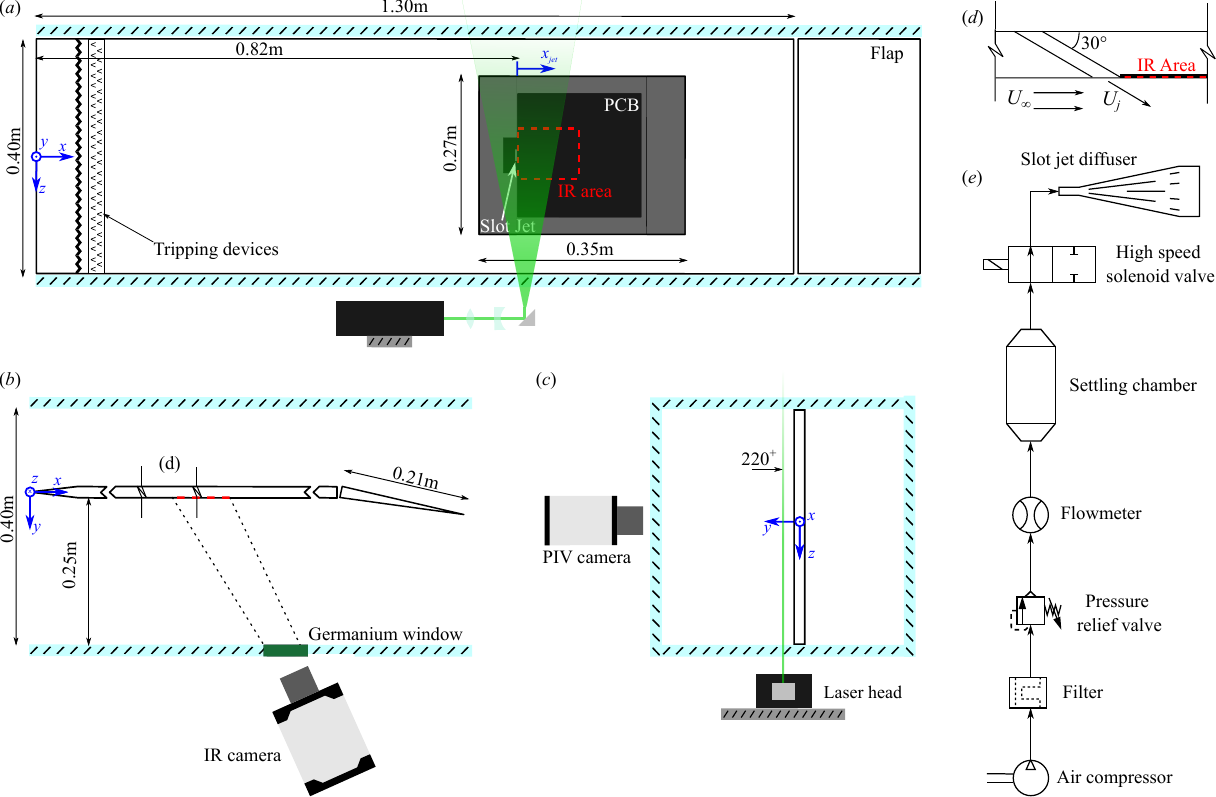}
    \caption{Schematic representation of the experimental setup in the wind tunnel: top (a), side (b) and back (c) views. The red dashed area \lcap{--}{red} indicates the IR-thermography measurement domain downstream of the actuator and the green plane \lcap{-}{green_laser} depicts the PIV illumination plane at $y^+=220$ from the wall. (d) Detailed view of the slot-jet flow injection. (d) Pneumatic system driving the jet secondary flow injection.}
    \label{fig:SetUp}
\end{figure*}

\subsection{Wind tunnel model and flow conditions \label{ss:WTandBLcond}}
The experimental campaign was carried out in the G\"ottingen-type wind tunnel at Universidad Carlos III de Madrid. The tunnel features a test chamber with a $0.4 \times 0.4~\mathrm{m}^2$ cross-section and a length of $1.5~\mathrm{m}$. Tests can be performed with a freestream velocity of up to $20~\mathrm{m/s}$, with a streamwise turbulence intensity below 1\% of the freestream velocity.

The turbulent boundary layer develops on a smooth poly-methyl methacrylate flat plate of $1.5~\mathrm{m}$ length and $20~\mathrm{mm}$ thickness, spanning the entire width of the test section (see figure~\ref{fig:SetUp}(a,b)). The flat plate is installed so that the flow develops under zero pressure gradient conditions (ZPG), which has been carefully checked by measuring the pressure coefficient profile. The plate is equipped with a spanwise aligned slot-jet actuator and with a wall heat flux sensor, both flush-mounted by means of inserts produced with additive manufacturing (see figure \ref{fig:SetUp}(a)).

The flat plate features a wedge-shaped leading edge. The location of the leading edge stagnation point is controlled by a $210~\mathrm{mm}$ long adjustable trailing-edge flap, deflected by approximately $10^\circ$. The boundary layer is tripped close to the leading edge with $2.5~\mathrm{mm}$ height zig-zag turbulators in combination with a $10~\mathrm{mm}$ wide DYMO\textsuperscript{\textregistered} tape (with the embossed letter ‘V’ pointing in the flow direction), eliminating coherent structures formed after the zig-zag turbulator strip and ensuring that the turbulent boundary layer at the measurement locations is not reminiscent of tripping effects \citep{sanmiguel2017diagnostic}.

A reference system with origin on the mid-section of the leading edge of the flat plate is considered. The velocity components along the streamwise $(x)$, wall-normal $(y)$ and spanwise $(z)$ directions are denoted by $U$, $V$ and $W$, respectively. Over-lined symbols refer to mean quantities (e.g. $\overline{U}$).
Wall-scaled variables are defined in terms of the local friction velocity, $u_\tau$, and kinematic viscosity, $\nu$, and are denoted by a superscript ‘+’. Other quantities used throughout the paper are the free-stream velocity, $U_\infty$, the actuator width, $\lambda$, and the boundary-layer thickness, $\delta$. 

The experiments are carried out for a freestream velocity of $U_\infty \approx 12~\mathrm{m/s}$, which correspond to a momentum thickness $\theta$ and friction-based Reynolds number $Re_\theta \approx 2000$ and $Re_\tau \approx 800$, respectively, at the jet location. There are several factors affecting the selection of the freestream velocity in this experiment: turbulent boundary layer development, velocity ratio with the cross-flow jet, or even infrastructure limitations. The chosen $U_\infty \approx 12~\mathrm{m/s}$ is a compromise between the available infrastructure capabilities and the development of a well-behaved turbulent boundary layer within the test region. Additionally, a suitable velocity ratio is assured so that the influence of the actuation is large enough to cause an effect that can be measured on the heat transfer sensor.

The TBL profile was measured with Particle Image Velocimetry (PIV). A dual cavity Nd:Yag Quantel Evergreen laser ($200~\mathrm{mJ/pulse}$ at $15~\mathrm{Hz}$) and a set of cylindrical and spherical lenses are employed to illuminate the DiEthyl-Hexyl-Sebacate seeding particles of $\approx 1~\mu\mathrm{m}$. An Andor sCMOS camera, equipped with a lens with $100~\mathrm{mm}$ focal length and using a focal ratio $f/\# = 11$, is used to image the flow with a resolution of approximately $51.7~\mathrm{pixels/mm}$. The eigenbackground removal procedure proposed by \citet{Mendez2017pod-piv} is used to remove the background. Statistics are computed using Ensemble Particle Tracking Velocimetry \citep{cowen1997hybrid}, with polynomial filtering of the particle clouds to improve the spatial resolution \citep{aguera2016EPTV}. The final averaging bin is $160\times3~\mathrm{pixels}$. 

The parameters of the reference TBL at the slot-jet streamwise location are computed by fitting the experimental data on the composite profile proposed by \citet{chauhan2009}. The reference TBL parameters are attached in table \ref{tab:TBL}, including the friction-based Reynolds number $Re_\tau$, the Reynolds number based on momentum thickness $Re_\theta$, the displacement and momentum thickness $\delta^*$ and $\theta$, the shape factor $H_{12}=\delta^*/\theta$ and the friction velocity $u_\tau$.

\begin{table}[t]
\caption{Parameters of the reference turbulent boundary layer at the slot-jet streamwise location $x_{jet}$.}
    \centering
        \begin{tabular}{>{\centering}p{0.125\linewidth-2\tabcolsep}
                >{\centering}p{0.125\linewidth-2\tabcolsep}
                >{\centering}p{0.125\linewidth-2\tabcolsep}
                >{\centering}p{0.125\linewidth-2\tabcolsep}
                >{\centering}p{0.125\linewidth-2\tabcolsep}
                >{\centering}p{0.125\linewidth-2\tabcolsep}
                >{\centering}p{0.125\linewidth-2\tabcolsep}
                >{\centering\arraybackslash}p{0.125\linewidth-2\tabcolsep}}
        \toprule
        $Re_\tau$ & $Re_\theta$ & $H_{12}$ & $\delta^*/\delta_{99} $ & $\theta/\delta_{99}$ & $\delta_{99}$ [mm] & $u_\tau$ [m/s] & $U_\infty$ [m/s]\\
        \midrule
        816 & 1948 & 1.35 & 0.144 & 0.107 & 23.2 & 0.54 & 12.2\\
        \bottomrule
        \end{tabular}
    \label{tab:TBL}
\end{table}

Uncertainties of the turbulent boundary layer parameters are computed with the tool proposed by \citet{Castellanos2021PIVuncertainty} for PIV/EPTV measurements. From this, the uncertainty of the displacement thickness $\delta^*$ and boundary layer thickness $\delta$ (approximated as $\delta \approx \delta_{99}$, such that $U(y=\delta_{99})=0.99U_\infty$) are found to be below $\pm 2\%$. The estimation of the freestream and local friction velocity has an uncertainty below $1\%$, with the position of the wall being estimated with an error below $\Delta y^+ = \pm 5$. 

The PIV data are also used to assess whether the TBL can be considered well-behaved. We adopt the method proposed by \citet{sanmiguel2017diagnostic} based on the diagnostic plot. The parameters are reported in table \ref{tab:diagnostic}, showing good agreement with DNS simulation from \cite{jimenez2010}.

\begin{table}
\caption{Diagnostic plot parameters of the non-actuated TBL at $x=x_{jet}$ and the DNS data from \citet{jimenez2010}. Error with respect to the method by \citet{sanmiguel2017diagnostic} between brackets.}\label{tab:diagnostic}
\begin{tabular*}{\tblwidth}{@{}LLCC@{}}
\toprule
Case & $Re_\theta$ & $\alpha$ & $\beta$ \\ \midrule
Reference TBL & 2015 & 0.295 ($+2.1\%$) & 0.265 ($+4.7\%$) \\
\citet{jimenez2010} & 1978 & 0.291 ($+0.3\%$) & 0.257 ($+1.5\%$) \\
\bottomrule
\end{tabular*}
\end{table}

\subsection{Actuation system \label{ss:actuator}}

The actuator is a slot-jet placed at a distance $x_{jet} = 815~\mathrm{mm}$ from the leading edge. From now on, the symbols $\hat{x}=(x-x_{jet})/\delta$ and $\hat{z}=z/\delta$ are used to refer to the freestream coordinate with origin in the jet and the spanwise coordinate, both normalised with the boundary layer thickness at the jet location.

The actuator shape is a slot of $\lambda = 25~\mathrm{mm}$ length and $w = 1~\mathrm{mm}$ width aligned with the spanwise direction. The spanwise orientation of the slot is preferred to investigate the two-dimensional effect of the jet actuation without the three-dimensional features coming from the horseshoe vortex generated at both sides of the jet.
The air injection in the TBL is performed at a $30^\circ$ angle with respect to the wall as shown in figure~\ref{fig:SetUp}(c). This inclination is suited to energize the lower region of the boundary layer, since the injected flow remains attached to the wall \citep{Steinfurth2021pulsedjet}. Additionally, this is a standard configuration for industrial applications, such as film cooling \citep{coletti2013}. A slot-jet diffuser was designed to ensure a smooth transition from the circular pneumatic line to the nozzle exit section. The diffuser total length of $70~\mathrm{mm}$ is divided in an inlet for a rounded connection of inner diameter $4~\mathrm{mm}$, an enlargement section of $50~\mathrm{mm}$ length and $15^\circ$ expansion angle. The enlargement section is equipped with vanes to improve the flow uniformity across the slot width and reduce the pressure losses within the diffuser. The design has been carried out with the support of numerical simulations.  

The pneumatic system used to feed the jet is sketched in figure~\ref{fig:SetUp}(d). Compressed air is filtered and regulated to a constant absolute pressure of $P_{jet} = 3.0 (\pm0.005)~\mathrm{bar}$ by means of a pressure-relief valve. The mass flow rate passing through the system is monitored by an Alicat Scientific\texttrademark~\textit{M-100SLPM} high-accuracy mass-flow meter, which is also employed to track the value of the absolute pressure and temperature in the pneumatic line. The jet pulsation is enabled by a SMC\textsuperscript{\textregistered}~\textit{SX-11DJ} high-speed solenoid valve which provides ON/OFF control at a frequency of up to $350~\mathrm{Hz}$. The solenoid valve is actuated by a $24~\mathrm{V}$ periodic square signal with a characteristic carrier frequency $f$ and duty cycle $DC$, defined as the percentage of the period $T$ in which the signal is set at high level. The square-wave signal driving the actuation is generated through MATLAB-Simulink\textsuperscript{\textregistered} and transferred to the valve via an Arduino\textsuperscript{\textregistered}~\textit{Uno} microcontroller. 
A characterisation of the slot-jet, based on hot-wire anemometry (HWA), concluded that the frequency response of the flow exiting the nozzle coincides with the desired actuation frequency for the whole actuation range. Moreover, the characterisation corroborates the uniformity of the flow injection through the slot-shaped nozzle after the diffuser with a velocity ratio $U_{jet}/U_\infty = 0.7$.

Regarding the duty cycle, HWA confirms the reduction in expected flow rate with increasing frequency. The SMC\textsuperscript{\textregistered}~\textit{SX-11DJ} solenoid valve is certified with a small difference of $0.05~\mathrm{ms}$ between the opening and closing lapse time ($\tau_{open} = 0.45~\mathrm{ms}$ and $\tau_{close} = 0.40~\mathrm{ms}$, respectively), that translates in a progressive reduction of effective injection time as frequency increases. The volumetric flow sensor measures this hardware limitation, which remains consistent for the whole actuation range, and it is taken into account in the result discussion. 

\subsection{Infrared thermography measurements \label{ss:IR}}

The convective heat transfer rate achieved by the system is assessed through the measurement of wall heat fluxes. Infrared thermography is used to obtain temperature maps of a heated-thin-foil heat-flux sensor as described by \citet{astarita2012infrared}. The chosen heat-flux sensor is a Printed Circuit Board (PCB), similar to those used by \citet{torre2018HTF}. The PCB is chemically bonded to an additively-manufactured insert made of polyethylene terephthalate glycol-modified (PETG) and flush-mounted into the flat plate (see figure \ref{fig:SetUp}(a,b)). The PCB features a copper track with a thickness of $5~\mathrm{\mu m}$, a pitch of $2~\mathrm{mm}$ and a gap of $0.2~\mathrm{mm}$. Thermal gradients across the thickness of the heat-flux sensor can be considered negligible \citep{astarita2012infrared}, since the Biot number ($\mathrm{Bi} = h t/k$, where $h$ is the convective heat transfer coefficient, $k$ the foil thermal conductivity coefficient and $t$ thickness, respectively) is sufficiently small $(\mathrm{Bi} \approx 0.003)$. The heat-flux sensor is placed just downstream of the slot-jet actuator. 

The PCB has a nominal resistance of $10~\Omega$, a circuit area $(A_{\mathrm{PCB}})$ of $200\times 200~\mathrm{mm}^2$ and a thickness of $0.8~\mathrm{mm}$. A constant heat-flux $q''_{j}$ is provided by Joule effect, achieved through a stabilised power supply providing a constant current $I \approx 2\mathrm{A}$ and voltage $V \approx 16.5~\mathrm{V}$ to the circuit ($q''_{j} = V I / A_{\mathrm{PCB}}$). The convective heat transfer coefficient $h$ is computed by posing a steady-state energy balance, modelling the PCB as a heated-thin-foil sensor \citep{astarita2012infrared}, and is expressed in non-dimensional form through the Nusselt number ($Nu = h \delta/k_{\mathrm{air}}$, where $k_{\mathrm{air}}$ is the thermal conductivity of air).
\begin{equation}
	h = \frac{ q''_{j} - q''_{r} - q''_{k} - q''_{b} }{ T_{w} - T_{aw} } ,
	\label{eq:heatedthinfoil}
\end{equation}
where $T_{w}$ is the surface temperature of the PCB, $T_{aw}$ is the adiabatic wall temperature, $q''_{r}$ is the radiation heat flux, $q''_{k}$ is the tangential conduction heat flux through the PCB, and $q''_{b}$ is the heat flux conducted through the thin air layer and the PETG substrate on the back side of the heated-thin-foil sensor. Note that the heat losses on the back side of the PCB are minimised by a $2~\mathrm{mm}$ gap between the PCB and the PTEG-insert. Natural convection on the front side of the PCB is negligible since $\mathrm{Gr/Re}^2<< 1$ ($\mathrm{Gr} = {g\beta_v(T_{w}-T_{aw}) L^3}/{\nu^2}$ is the Grashoff number, where $\beta_v$ is the volumetric thermal expansion coefficient and $g$ is the acceleration of gravity). Radiative heat flux to the environment is estimated assuming that the environment behaves as a black body at a temperature equal to that of the freestream, so that $q''_{r} = \sigma \varepsilon (T_w^4 - T_\infty^4)$ (where $\sigma$ is the Boltzmann constant and $\varepsilon$ is the emissivity of the PCB surface). The PCB layout is compliant with the criteria proposed by \citet{torre2018HTF} for heated-thin-foil sensors, with losses to tangential condition, $q''_{k} = kt \left( \frac{\partial^2T}{\partial x^2} + \frac{\partial^2T}{\partial y^2} \right)$, below $0.5\%$. The interested reader is referred to \citet{astarita2012infrared} for a profound description of more sophisticated methods which distinguish between tangential heat flux along the direction parallel to the PCB tracks and normal to the tracks. Lastly, $q''_{b}$ is estimated by considering the conduction through the air gap and the PETG substrate, $q''_{b} = \left(t_{\mathrm{air}}/k_{\mathrm{air}}+t_{\mathrm{PETG}}/k_{\mathrm{PETG}}+1/h_b\right)^{-1}(T_{w}-T_{aw})$, being $t_m$ and $k_m$ the thickness and thermal conductivity of material $m$, and $h_b$ the convective heat transfer coefficient on the back side of the plate assumed to be equal to the average turbulent heat transfer on the non-actuated TBL.

Temperature measurements of the heated-thin-foil sensor are performed with a Infratec 8820 IR camera ($640 \times 512\mathrm{pix}$ Mercury-Cadmium-Telluride detector and Noise Equivalent Temperature Difference $< 25\mathrm{mK}$), capturing images at a frequency of $50\mathrm{Hz}$ with a spatial resolution of $3.8\mathrm{pixels/mm}$. To ensure the accuracy of IR temperature measurements the PCB is coated with a thin layer of high-emissivity paint ($\varepsilon = 0.95$). The adiabatic wall temperature ($T_{aw}$) is computed from a set of $500$ \textit{cold} images acquired with no heating to the PCB ( $q_j''= 0$); the wall temperature ($T_{w}$) derives from a set of $1000$ \textit{hot} images acquired with heating to the PCB activated. The uncertainties associated with the experimental measurements are calculated by a Monte Carlo simulation \citep{minkina2009infrared}, assuming statistically-uncorrelated errors and using the uncertainty values reported in table~\ref{tab:uncertainties}.
The uncertainty on the Nusselt number is estimated to be lower than $\pm 4\%$. Notwithstanding, the IR thermography measurements were conducted three times on different days to ensure the repeatability of the phenomena.

\begin{table}
\centering
\caption{Uncertainty Analysis on Nusselt number calculation} \label{tab:uncertainties}
\begin{tabular*}{\tblwidth}{@{}LLL@{}}
\toprule
    Parameter               & Uncertainty   & Typical Value \\ \midrule
    $T_w$                   & 0.1 K         & 310 [K]  \\
    $T_{aw}$                & 0.1 K         & 300 [K] \\
    $T_\infty$              & 0.1 K         & 305 [K] \\
    $V$                     & 0.2\%         & 17 [V] \\
    $I$                     & 0.2\%         & 2 [A] \\
    $\varepsilon$           & 2\%           & 0.95 \\
    $A$                     & 0.1\%         & 441 [cm$^2$] \\
    $k_{{\mathrm{air}}}$    & 1\%           & 0.0265 [W/(m K)] \\
    $k_{{\mathrm{PETG}}}$   & 5\%           & 0.200 [W/(m K)] \\
    $t_{{\mathrm{air}}}$    & 1\%           & 2.0 [mm] \\
    $t_{{\mathrm{PETG}}}$   & 1\%           & 8.0 [mm] \\
    $\delta$                & 1\%           & 23 [mm] \\
    $U_\infty$              & 3\%           & 12 m/s \\
    $q''_k$                 & 10\%          & 0.5 [mW/m$^2$] \\ 
    $h_b$                   & 4\%           & 55 [W/(m$^2$ K)] \\ 
\bottomrule
\end{tabular*}
\end{table}


\subsection{Velocity measurements \label{ss:PIV}}
Velocity field measurements are performed to complement the heat transfer distribution from IR thermography. Two-component Particle Image Velocimetry (PIV) measures streamwise and spanwise velocity fields in a wall-parallel plane at a distance of $y^+ \approx 220$ from the wall, namely $y\approx 6.3\mathrm{mm}$ ($y/\delta = 0.27$), as shown schematically in figure \ref{fig:SetUp}(c). DiEthyl-Hexyl-Sebacate seeding particles of $\approx 1~\mu\mathrm{m}$ size are illuminated by a dual cavity Nd:Yag Quantel Evergreen laser ($200~\mathrm{mJ/pulse}$ at $15~\mathrm{Hz}$). A set of cylindrical and spherical lenses is used to shape a laser beam in a plane with a thickness of approximately $1~\mathrm{mm}$ in the measurement region. The imaging is achieved by an Andor sCMOS camera, equipped with a lens with $50~\mathrm{mm}$ focal length and using a focal ratio $f/\# = 11$. The field of view is cropped over the region of interest just downstream the slot, covering $140\times60~\mathrm{mm}^2$ ($0 \leq \hat{x} \leq 0.140~\mathrm{m}$ and $|z|\leq 0.03~\mathrm{m}$), which results in a resolution of approximately $17~\mathrm{pixels/mm}$. 

Velocity data is derived from the ensemble average of $3000$ image pairs acquired at $15\mathrm{Hz}$. The images are pre-processed to remove the background through the eigenbackground removal procedure \citep{Mendez2017pod-piv}. The herein-employed PIV processing software was implemented by the Experimental Thermo-Fluid-Dynamics Group of the University of Naples Federico II, with a multi-pass cross-correlation algorithm with window deformation \citep{soria1996piv,scarano2001iterativeimgdef} and advanced interpolation schemes and weighting windows to improve the spatial resolution and precision of the process \citep{Astarita2006PIV,Astarita2007PIV}. The final interrogation window of the PIV process has size of $48\times48~\mathrm{pixels}^2$ and 75\% of overlap, resulting in $1.4~\mathrm{vector/mm}$.


\section{Results and discussion on heat transfer \label{s:Results}}

\begin{figure*}[]
\centering
    \includegraphics*[width=0.95\linewidth]{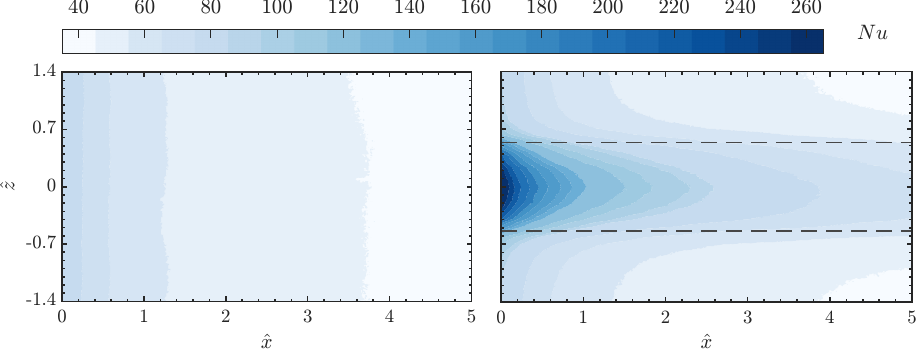}
\caption{\label{fig:Numaps} Nusselt number distribution downstream of the slot jet for the non-actuated TBL (left) and the steady jet actuation (right). The limits of the slot are highlighted \lcap{--}{black}.} 
\end{figure*}

Infrared thermography measurements are performed directly downstream of the actuation on a region comprising $5\delta \times 2.6\lambda$.
The Nusselt number distribution is characterised in terms of two reference conditions: the non-actuated TBL and the steady-jet actuation. The spatial distributions for both cases are shown in figure~\ref{fig:Numaps}. The Nusselt number distribution for the non-actuated case (denoted by subscript $0$) agrees with the literature, describing a progressive decay of $Nu \sim x^{-0.2}$ for a conventional TBL \citep{Lienhard2020} and a range of values that agrees with similar experiments in the same facility \citep{mallor2018cubes}. Regarding the steady-jet actuation (denoted by subscript $sj$), the Nusselt number distribution exhibits a symmetric pattern with a strong convective heat transfer enhancement in the vicinity of the air injection. The effect of the jet is persistent even $5\delta$ downstream of the actuator; however, the enhancement is very localised in the spanwise direction, mainly affecting the area downstream of the jet. Consequently, spatial averaged quantities (namely over-lined symbols, e.g. $\overline{Nu}$) are computed in the region comprised in $0\leq \hat{x} \leq 5$ and $-\lambda/2 \leq z \leq -\lambda/2$: 
\begin{equation} \label{eq:Num}
    \overline{Nu} = \frac{1}{\lambda} \int_{-\lambda/2}^{\lambda/2} \int_{0}^{5} Nu(x,z) \,d\hat{x}\,dz
\end{equation}

A summary of the main non-actuated case and steady-jet actuation features is included in table~\ref{tab:steady-jet} for reference, including the relative enhancement $E=\overline{Nu}_{\mathrm{sj}}/\overline{Nu}_0 -1 $.

\begin{table}[t] 
    \centering
    \caption{Reference performance indicators of steady-jet and non-actuated TBL.}
    \label{tab:steady-jet}
        \begin{tabular}{>{\centering}p{0.2\linewidth-2\tabcolsep}
                >{\centering}p{0.245\linewidth-2\tabcolsep}
                >{\centering}p{0.185\linewidth-2\tabcolsep}
                >{\centering}p{0.185\linewidth-2\tabcolsep}
                >{\centering\arraybackslash}p{0.185\linewidth-2\tabcolsep}}
        \toprule
        $\dot{V}_{\mathrm{sj}}$ [L/s] & $U_{jet,sj}$ [m/s] & $\overline{Nu}_{0}$ & $\overline{Nu}_{\mathrm{sj}}$ & $E$ \\
        \midrule
        0.22 & 8.77 & 50.4 & 91.9 & 0.82\\
        \bottomrule
        \end{tabular}
\end{table}

Concerning the actuated flow case, the parametric study covers a wide space of the two actuation parameters: carrier actuation frequency $f$ from $25\mathrm{Hz}$ to $300\mathrm{Hz}$ in steps of $25\mathrm{Hz}$, and duty cycle $DC$ limited to $25\%$, $50\%$ and $75\%$. The combination of frequencies and duty cycles gathers $36$ control laws together with the steady-jet actuation and the non-actuated TBL.

\subsection{Wall distribution of the Nusselt number}
\begin{figure*}[t] 
    \centering\
    \includegraphics*[width=0.95\linewidth]{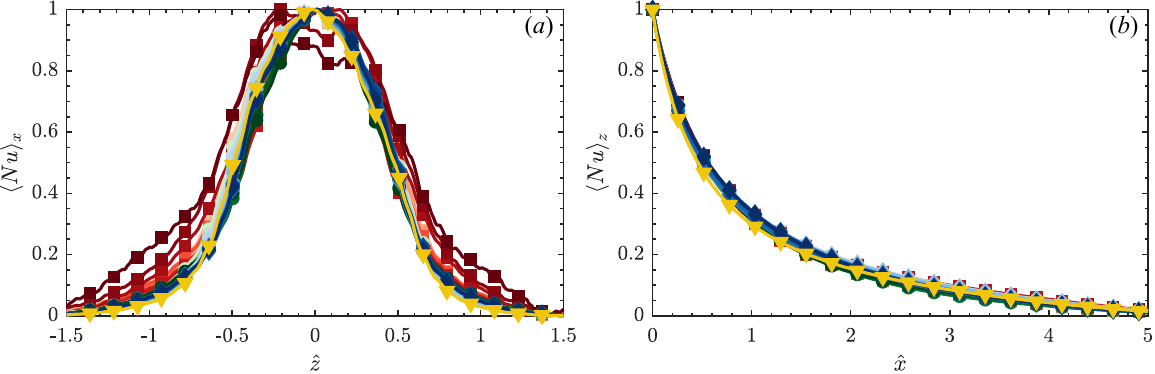}
    \caption{Nusselt profiles averaged along the streamwise (a) and spanwise (b) direction for both the steady-jet actuation and all the pulsed-jet configurations. The profiles are scaled with a \textit{min-max} normalisation. Duty cycle is depicted by marker and color, following: \sy{Blue1}{d*}75\% DC, \sy{Green1}{o*}50\% DC, \sy{Red1}{s*}25\% DC and \sy{Yellow1}{t*} steady. Colours range from light to dark tone for increasing actuation frequency.}
    \label{fig:similarity}
\end{figure*}

For the sake of clarity and conciseness, Nusselt number results are reported in the spatial average form, as described above. The wall distribution of the different cases, as shown for the two reference flow conditions in figure~\ref{fig:Numaps}, does not suggest any direct outcome at first sight, so post-processing of the data was required to obtain conclusions. It is to be noted that, all the tested control laws present a similar spatial distribution in terms of Nusselt number. 
To stress out this point, the spatial similarity among the tested control actions is analysed in Figure~\ref{fig:similarity}, showing the Nusselt profiles averaged through the streamwise and spanwise directions for both the steady-jet actuation and all the pulsed-jet configurations. The profiles are scaled with a \textit{min-max} normalisation to focus on the shape of the induced heat transfer rather than on the magnitude.
The similarity in the heat transfer distribution from the different control actions suggests that both the effects of $f$ and $DC$ affect mostly the intensity of the actuation while its pattern changes very weakly.

According to \citet{Sau2010optJICF}, there is still no consensus on the physical mechanisms involved in the optimisation of jet penetration and on a proper scaling law for the JICF. Some studies \citep[e.g.][]{MCLOSKEY2002,shapiro2006optimization} rely on  the ``formation number'', proposed by \citet{gharib1998scalingVR} for vortex ring formation in a quiescent medium, to express the optimal pulsation condition. \citet{Johari2006scaling}, in his attempt to identify a universal scaling for JICF, demonstrated that ``stroke ratio'' and ``duty cycle'' are the only parameters describing pulsed JICF. This approach is however only valid for quiescent flow conditions, as confirmed by \citet{Sau2008dynamicsVRICF}. Later, \citet{Sau2010optJICF} proposed a regime map based on the ``ring velocity ratio'' and the ``stroke ratio'' gathering the effects of the pulsation frequency, duty cycle, modulation, and pulsed energy. They demonstrate that many of the experimental investigations for the optimal pulsation of JICF up to that day could be interpreted in terms of their vortex rings, and that the optimal experimental conditions are seen to collapse on the same optimal curve.
Despite the extensive investigations for the proper scaling of JICF dynamics, the common choice to express the optimal pulsation conditions is the Strouhal number, $St$; however, the literature on pulsed JICF reports a wide range of optimal values for $St$ that commonly depends on other factors such as flow conditions, type of modulation or the apparatus itself \citep{MCLOSKEY2002}. Based on the literature investigating JICF, there are several possibilities for defining $St$. The most accepted trend relies on the jet injection velocity and the diameter of the jet, while several past studies prefer to assess the performance of the actuation on the boundary layer, hence using the boundary layer thickness and the freestream velocity as scaling parameters. Since the definition of the Strouhal number mainly depends on the phenomena of focus, in this study $St$ is customarily defined as,
\begin{equation}\label{eq:St}
    St = \frac{f \cdot \delta}{c},
\end{equation}
being $c$ the convection/advection velocity of large-scale turbulent structures moving within the TBL. The advection velocity value is estimated based on literature \citep{Krogstad1998convvel,jimenez2004lsturb}, being $c \approx 10\cdot u_\tau = 5.2 \mathrm{m/s}$. 
While this choice might seem arbitrary, we hypothesize that the convective heat transfer enhancement is mainly to be ascribed to the production of coherent structures in the buffer layer. This decision is founded on the extensive literature on flow structure upon the actuation of a JICF, coinciding in coherent structures like vortex rings, hairpin vortices, spanwise rollers or trailing columns that commonly develop in the lower region of the boundary layer. Hence, according to this definition of the Strouhal number, the frequency corresponding to $St = 1$ can be understood as the one exciting the turbulent coherent motions with a size of the order of $\delta$ that move within the TBL with a velocity $c$. 

\begin{figure}[t] 
    \centering\
    \includegraphics*[width=0.99\linewidth]{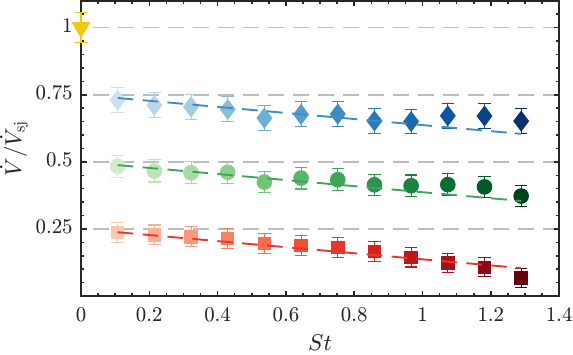}
    \caption{Pulsed jet volumetric flow rate $\dot{V}$ for the tested actuation frequencies and duty cycles. Results are normalised with respect to the volumetric flow rate of the steady jet, $\dot{V}_{\mathrm{sj}}$. Duty cycle is depicted by marker and color, following: \sy{Blue1}{d*}75\% DC, \sy{Green1}{o*}50\% DC, \sy{Red1}{s*}25\% DC and \sy{Yellow1}{t*} steady. Colours range from light to dark tone for increasing actuation frequency. Dashed lines represent the linear model described in Eq.\eqref{eq:Q/Qst}}.
    \label{fig:mdot}
\end{figure}

The duty cycle is a parameter of the actuation directly related to the injected volumetric flow rate from the jet. The three tested cases are enough to understand its influence on convective heat transfer. Figure~\ref{fig:mdot} shows the value of the volumetric flow rate for each of the tested cases with respect to the steady jet. For the three tested $DC$ values, a monotonic decay with $St$ is observed, which is directly related to the technological limitations imposed by the valves (described in Section~\ref{ss:actuator}). Based on the experimental data, the effective duty cycle $DC^*$ is defined as, 
\begin{align}\label{eq:effDC}
\begin{split}
    DC^* & =\frac{\tau}{T}\\
         & = \frac{DC/f-\tau_{open}+\tau_{close}}{1/f}\\
         & = DC -f \left(\tau_{open}-\tau_{close}\right) \\
         & = DC -St \left(\tau_{open}-\tau_{close}\right)\frac{c}{\delta_{jet}} \\
         & = DC - \sigma \cdot St,
\end{split}
\end{align}
being $\tau$ the pulse width, $T = 1/f$ the actuation period, and $\tau_{open}$ and $\tau_{close}$ the time lapse for the valve to open and close, as defined in Section~\ref{ss:actuator}. The constant $\sigma = 0.1164$ is the normalised time difference of $0.05\mathrm{ms}$ between opening and closing lapses, namely $\sigma = (\tau_{open}-\tau_{close})c/\delta_{jet}$. This linear decay fits with the data reported in figure~\ref{fig:mdot} which means that the inconsistency between $DC$ and $\dot{V}/\dot{V}_{\mathrm{sj}}$ is solely related to technological constraints. In accordance with \citet{MCLOSKEY2002}, the performance of the pulsation of a jet in crossflow is directly dependent on the apparatus; however, the monotonic decay in $\dot{V}$ with $St$ is well characterised from equation~\ref{eq:effDC}, which is relevant for the following discussion along with the manuscript.

\begin{figure}[] 
    \centering\
    \includegraphics[width=0.99\linewidth]{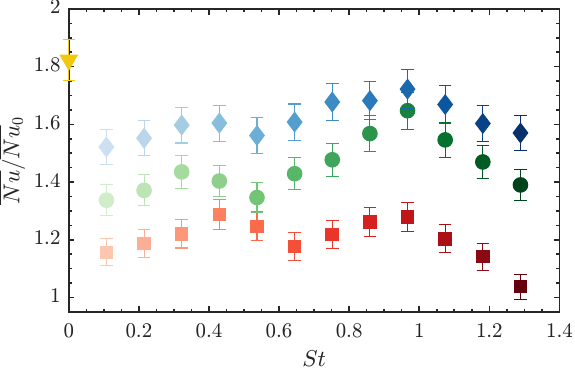}
    \caption{Averaged Nusselt number $\overline{Nu}$ downstream of the jet normalised with the non-actuated flow case $\overline{Nu}_0$. Refer to figure \ref{fig:mdot} for marker legend.}
    \label{fig:Num}
\end{figure}

The spatial average of the Nusselt number $\overline{Nu}$ for each control law is depicted in figure~\ref{fig:Num}. Values are referenced to that of the non-actuated flow $\overline{Nu}_0$. For a given duty cycle, the experimental results suggest the existence of two actuation frequencies at which the Nusselt number is maximised for a given $DC$. The local maximum is achieved at $St \approx 1$, while a second peak rises in the vicinity of $St \approx 1/3$. Based on the definition of the Strouhal number in equation~\ref{eq:St}, the maximum Nusselt number is achieved for the pulsation frequency that excites turbulent structures in the near-wall region; however, further analysis is required to confirm the physical phenomena behind this optimal pulsation. The frequency corresponds to $f=225\mathrm{Hz}$, which is very similar to the dominant shear layer frequency of the steady jet $f_{\mathrm{sj}}=220\mathrm{Hz}$ reported by \citet{MCLOSKEY2002} and \citet{shapiro2006optimization}. However, \citet{MCLOSKEY2002} describes that the highest penetration and amplitude are achieved at $1/2$, $1/3$ and $1/4$ of $f_{\mathrm{sj}}$. Notwithstanding that most of the literature dealing with pulsed jets in crossflow describe similar phenomena, there is a lack of consensus about which is the optimal $St$ value either for mixing or heat-transfer enhancement. Based on the common definition of the Strouhal for a rounded jet as $St_{jet} = D\cdot f/U_{jet}$ (being $D$ the jet diameter), it can be adjusted to the slot-jet application as $St_{jet} = \lambda \cdot f/U_{jet}$. The two peaks are found at $St_{jet} = 0.64$ and $0.21$, respectively. \citet{eroglu2001structure} found the optimal penetration in the range of $St_{jet}=0.14-0.42$, while for $St_{jet} > 0.7$ the behaviour decays rapidly. \citet{vermeulen1992mixing} investigated a cold jet with hot crossflow, concluding that the optimal mixing is achieved for $St_{jet} \approx 0.27$; however, they did not investigate above $St_{jet} > 0.4$ to provide further conclusions. Nonetheless, it is to be noted that all these contributions worked with high-velocity ratios ($1.5 \leq R \leq 4$) compared to the herein investigated $R = 0.73$, which translates into a lower control authority to alter the base flow and in the dominance of the boundary layer dynamics with respect to the jet dynamics.

\subsection{Performance and Interpretation of heat transfer enhancement}\label{s:enhanment_model}
The convective heat transfer enhancement with respect to the non-actuated boundary layer has been partially assessed from the results reported in figures~\ref{fig:Num}. However, some further aspects can be noted in the performance of the actuation laws. The duty cycle has a direct effect on the enhancement since it is directly related to the amount of momentum injected from the jet and hence the power added into the system. The observed linear trend with the $DC$ was thus a quite expected result. On the other hand, the effect of the pulsation frequency does not seem to have a direct trend.

In this section, a simplified model is proposed to decouple the effects of the duty cycle and frequency in heat transfer enhancement. The normalised heat transfer enhancement $E(f,DC)$ is defined as the relative improvement in the Nusselt number of the actuated case with respect to the non-actuated TBL:
\begin{equation}
    E(f,DC) = \frac{\overline{Nu}(f,DC)}{\overline{Nu}_0}-1,
    \label{eq:Edef}
\end{equation}

\noindent being $Nu(f,DC)$ the Nusselt number of the actuated configuration with frequency $f$ and duty cycle $DC$, respectively. The duty cycle is a parameter of the actuation directly related to the injected mass flow rate; however, the performance of the valves directly affects the net volumetric flow. This aspect is addressed in Section~\ref{s:Results}, describing the  effective duty cycle $DC^*$ in equation~\ref{eq:effDC}. Consequently, and based on the experimental data reported in figure~\ref{fig:mdot}, the first assumption is considered, 
\begin{equation}\label{eq:Q/Qst}
   \frac{\dot{V}}{\dot{V}_{\mathrm{sj}}} \approx DC^* = DC - \sigma \cdot St
\end{equation}

Given the dependency of experimental data on the previous approximation of the valve performance, the heat transfer enhancement can be then expressed as a function of the effective duty cycle and the actuation frequency, dividing the contribution of each parameter into two subfunctions,
\begin{equation}
    E(f,DC^*) = E_f(f) + E_{DC}(DC^*)
\end{equation}

Based on the alteration of the Nusselt number for different $DC$ values, it is straightforward to consider the relation of $Nu$ enhancement with $DC$ as linear:
\begin{equation}\label{eq:Edc1}
    E_{DC}(DC^*) = \kappa \cdot DC^*
\end{equation}
being $\kappa$ a constant of proportionality. Given the steady-jet case with $DC = DC^* = 1$ and without dependency on frequency, the value of the proportionality constant reads as follows:
\begin{equation}\label{eq:Edc2}
    E_{\mathrm{sj}} = E_{DC} = \kappa \quad \rightarrow \quad \kappa = \frac{\overline{Nu}_{\mathrm{sj}}}{\overline{Nu}_0}-1
\end{equation}

Introducing the empirical model for the effective duty cycle in equation~\ref{eq:Q/Qst}, a simplified, linear superposition model is obtained for the decoupled enhancement of $Nu$ from the duty cycle: 
\begin{equation}\label{eq:Edc}
E_{DC} = \left(\frac{\overline{Nu}_{\mathrm{sj}}}{\overline{Nu}_0}-1\right) \cdot \left(DC - \sigma \cdot St\right)
\end{equation}

\begin{figure}[] 
    \centering
    \includegraphics*[width=0.99\linewidth]{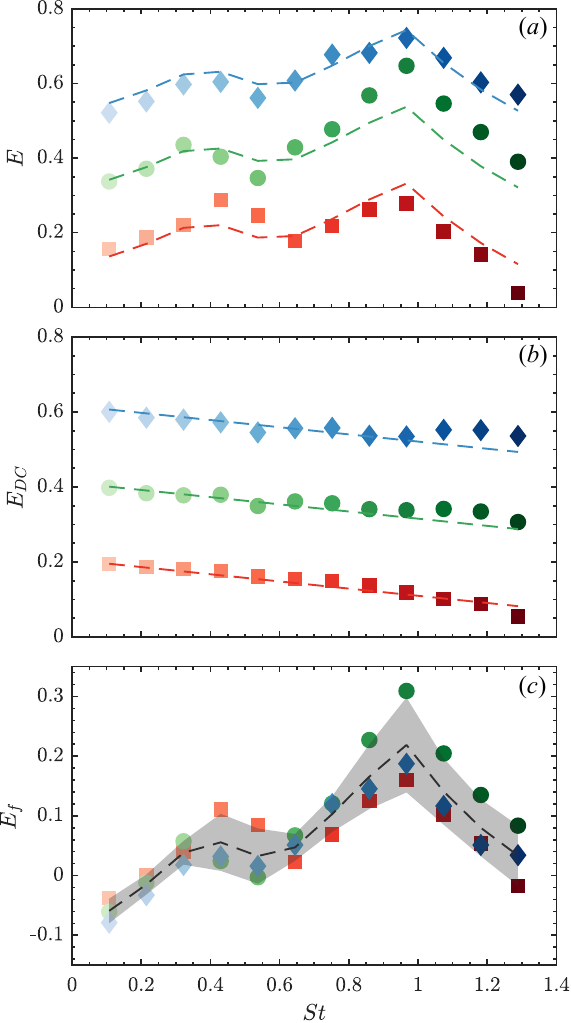}
    \caption{Decoupling of actuation parameters effect on convective heat transfer enhancement. 
    (a) Overall enhancement $E$ from equation~\ref{eq:Edef}; dashed lines \lcap{--}{black} represent the sum of $E_{DC}$ estimated from equation~\ref{eq:Edc} and the average $E_f$ depicted in (c).
    (b) Duty cycle decoupled contribution  $E_{DC}$; dashed lines \lcap{--}{black} represent the estimated $E_{DC}$ from equation~\ref{eq:Edc}.
    (c) Frequency decoupled enhancement $E_f$; dashed line \lcap{--}{black} represent the mean trend and the patch \sy{grey}{rec} highlights the confidence interval of the data.
    Refer to figure \ref{fig:mdot} for marker legend.}
    \label{fig:Enhancement}
\end{figure}

The validation with experimental data is depicted in figure~\ref{fig:Enhancement}(b), where the markers indicate the values obtained from the flow meter measurements ($\kappa\cdot \dot{V}/\dot{V}_{\mathrm{sj}}$). The decoupled effect of the duty cycle is in excellent agreement with equation~\ref{eq:Edc}. It is to be noted that, the dependency of this model on the Strouhal number (pulsation frequency) is solely related to the hardware limitations imposed by the valves, which is considered in the $\sigma$ value. Theoretically speaking, assuming no detrimental performance of the hardware involved in the experiment, the component $E_{DC}$ would be just dependent on $DC$, with $\sigma = 0$.

Since no information is available regarding the individual effect of the frequency, the estimation of $E_f$ is done by subtracting the experimental result for $E_{DC}$ component ($\kappa\cdot \dot{V}/\dot{V}_{\mathrm{sj}}$) from the overall $E$. The following mathematical derivation is used to decouple the data in figure~\ref{fig:Enhancement}(c),
\begin{align}\label{eq:Ef}
\begin{split}
      E_f & = E - E_{DC} \\
          & = \left(\frac{\overline{Nu}}{\overline{Nu}_0}-1\right) - \left(\frac{\overline{Nu}_{\mathrm{sj}}}{\overline{Nu}_0}-1\right) \cdot \left(DC - \sigma \cdot St\right)
\end{split}
\end{align}
Since the decoupling model was built under the assumption that $E_f$ is solely a function of the frequency, the contributions from the duty cycle embedded in $E$ and $E_f$ must cancel out so that the output of equation~\ref{eq:Ef} depends on the Strouhal number (and consequently on the actuation frequency). The data in figure~\ref{fig:Enhancement}(c) illustrate the dependency on $St$ with a certain confidence interval. The inconsistency among the different tested $DC$ values is expected from the experimental uncertainty. 

The simplified model is tested by reconstructing the overall enhancement $E$ in figure~\ref{fig:Enhancement}(a). The reconstructed data (dashed lines) fits reasonably well with the experimental data. The reconstruction is obtained from the empirical model in equation~\ref{eq:Edc} and the average $E_f$ function (dashed line figure~\ref{fig:Enhancement}(c)). These results conclude the possibility of decoupling the effect of the pulsation parameters $DC$ and $f$ when assessing the heat transfer enhancement in a turbulent boundary layer. The pulsation frequency is solely responsible for the existence of local maxima for the tested parametric space, while the duty cycle effect translates into a direct bias on the frequency response.

\begin{figure}[] 
    \centering
    \includegraphics*[width=0.99\linewidth]{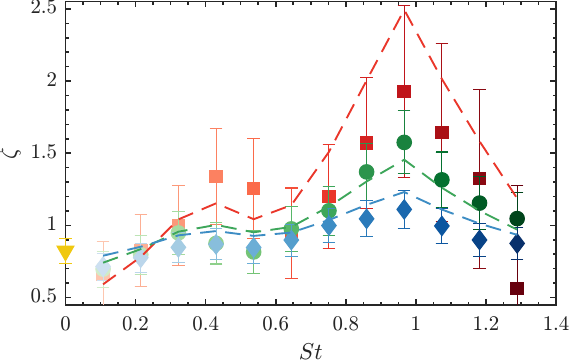}
    \caption{Performance indicator of the actuated jet for Nusselt number maximisation $\zeta = (\overline{Nu}/\overline{Nu}_0 -1) / (\dot{V}/\dot{V}_{\mathrm{sj}})$; dashed lines \lcap{--}{black} represent the estimated value from the proposed model. Refer to figure \ref{fig:mdot} for marker and colour legend.}
    \label{fig:eta}
\end{figure}

Based on the above-described effect of the duty cycle and the pulsation frequency on the heat-transfer enhancement, a performance analysis is required to determine the efficiency of the control. Despite the fact that the pulsed jet has a lower performance than the steady-jet case when referring to the maximisation of the Nusselt number (see figure~\ref{eq:Num}), a trade-off analysis is needed to accurately determine the performance of actuation including the cost of the actuation. For example, the actuation at $DC=75\%$ and $St=1$ provides a heat transfer enhancement close to that of the steady-jet while utilising $<3/4$ of the volumetric flow rate, which translates into $<3/4$ of the power requirement. To extend this thought to the rest of the control laws, a performance indicator $\zeta$ was defined as: 
\begin{equation}\label{eq:eta}
    \zeta = \frac{ \overline{Nu}/\overline{Nu}_0-1 }{ \dot{V}/\dot{V}_{\mathrm{sj}}},
\end{equation}
which compares the heat transfer enhancement with respect to the non-actuated flow with the volumetric flow rate injected through the jet. It must be remarked here that the actuation frequency and duty cycle do not affect the line pressure nor its velocity; consequently, this quantity $\zeta$ can be regarded as a ratio between a non-dimensional heat-transfer enhancement and a fair measure of the power output relative to the steady jet. As indicated in figure~\ref{fig:eta}, even if the overall enhancement is lower for actuated cases, the ``efficiency'' in providing that amount of enhancement is significantly higher than that of a steady jet. This is especially true for cases with actuation at $DC=25\%$, which at the ideal frequency present almost a $100\%$ increase in $\zeta$ with respect to the steady-jet case. These results pose a significant constraint to the decision of the control law; higher $DC$ and thus a higher injection of mass into the flow will provide a higher convective heat transfer, but at a lower $\zeta$ value. On the contrary, by maximising $\zeta$ and reducing the power consumption of the system, also reducing the injected mass, the convective heat transfer enhancement is weaker. Then, the selection of the actuation parameters will ultimately be driven by the requirements and constraints of the application. Furthermore, it is worth noting that the predicted values of $\zeta$ for heat transfer enhancement, based on the proposed model, lie within the uncertainty of the experimental data. The prediction is based on Eq.\eqref{eq:Edc} for the estimation of $E_{DC}$, and on the mean $E_f$ derived from Eq.\eqref{eq:Ef}, which is illustrated in figure~\ref{fig:Enhancement}(c) with a dashed-black line \lcap{--}{black}.

Even though tangential to the main focus of the paper on the fluid-dynamic mechanism behind the heat transfer enhancement, it is worth discussing the energy requirement of the actuation from a technological implementation standpoint. Since the pressure of the pumping line $\Delta p$ is fixed for the whole study, one could relate the energy requirement as the pumping power, namely $P_{\mathrm{jet}} = \dot{V}\Delta p$, which translates into approximately $45\mathrm{W}$ for the steady jet actuation and $10, 20, 30 \mathrm{W}$ for $DC = 0.25,0.5,0.75$, respectively. The fraction of this power that is effectively used to raise the convective heat transfer ($Q_{c} = A h \Delta T$, being $A$ the area on the region of interest used in this study downstream of the jet) compared to the unperturbed TBL ranges from 0.3\% to 1\%. This figure of merit concludes on the low technological performance of the proposed strategy, which could be considerably improved by reducing the pumping pressure, reducing the nozzle size to limit the volumetric flow rate or by distributing more actuators to homogenise the effect. We however emphasised that it is not the aim of this manuscript to discuss the energy efficiency of the control. We leave these considerations to future studies with the focus on a implementation of this solution from an engineering perspective.

%
\section{Flow dynamics assessment}\label{s:pivresults}
The heat transfer measurements are complemented with an analysis of the flow topology upon the action of the pulsed slot jet to further comprehend the flow structures and mechanisms involved in the heat exchange process at the wall. PIV measurements are performed in a wall-parallel plane at the outer edge of the logarithmic layer close to the transition to the wake of the TBL profile, namely $y^+=220$. At this distance from the wall, the bulk streamwise velocity of the unperturbed TBL is $U^* = U(y^+ = 220) = 9.85\mathrm{m/s} \approx 0.8U_\infty$. The PIV measurements focus on two pulsation frequencies $f = 150\mathrm{Hz}$ and $f = 225\mathrm{Hz}$ (namely, $St = 0.64$ and $0.97$), which are representative of a low- and high-performance actuation, respectively, according to the value of $\zeta$. 

\begin{figure*}[t] 
    \centering
    \includegraphics[width=0.9\linewidth]{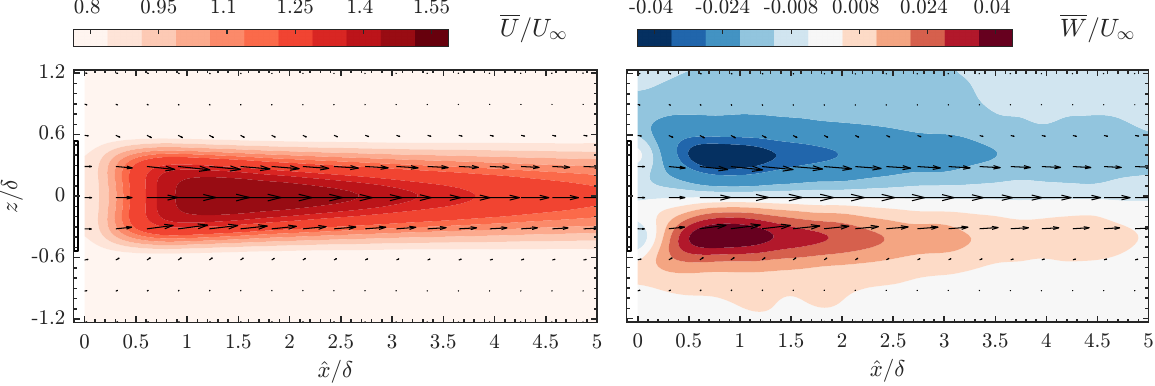}
    \caption{Streamwise $U$ and spanwise $W$ averaged flow fields of the steady-jet actuation on a wall-parallel plane at $y^+ = 220$. The vector field depicts $\overline{U}-U^*$ and $\overline{W}$.}
    \label{fig:PIV_ref}
\end{figure*}

Figure~\ref{fig:PIV_ref} illustrates the time-average streamwise $\overline{U}$ and spanwise $\overline{W}$ velocity components for the steady jet actuation. The superimposed vector field depicts the mean velocity field after removing the bulk streamwise velocity, namely $\overline{U}-U^*$. The air injection considerably accelerates the flow downstream of the slot in a narrow spanwise band. The local mean flow acceleration favours the entrainment of fluid from the surrounding, inducing a symmetric spanwise motion of flow towards the centre line. Furthermore, the flow topology resembles the observed distribution of the average Nusselt number (figure \ref{fig:Numaps}), with a prolonged region of persistent heat transfer enhancement favoured by the entrained flow. 

The recent numerical and experimental investigation by \citet{puzu2019jet} studied the flow topology and heat transfer enhancement on a TBL altered by a steady, rounded JICF with a similar velocity ratio as the herein considered. They describe a bifurcation of the mainstream around the jet exit that merges back after a certain distance downstream. The flow diversion is characterised by the induction of a counter-rotating vortex pair at both sides of the jet that promotes the mixing of the freestream and the injected fluid, especially upon the merging area downstream of the actuation. However, this study differs from the herein presented in three relevant aspects: the flow pulsation, the slot shape of the jet, and the inclination of the injection plane.

\begin{figure*} 
    \centering
    \includegraphics[width=0.9\linewidth]{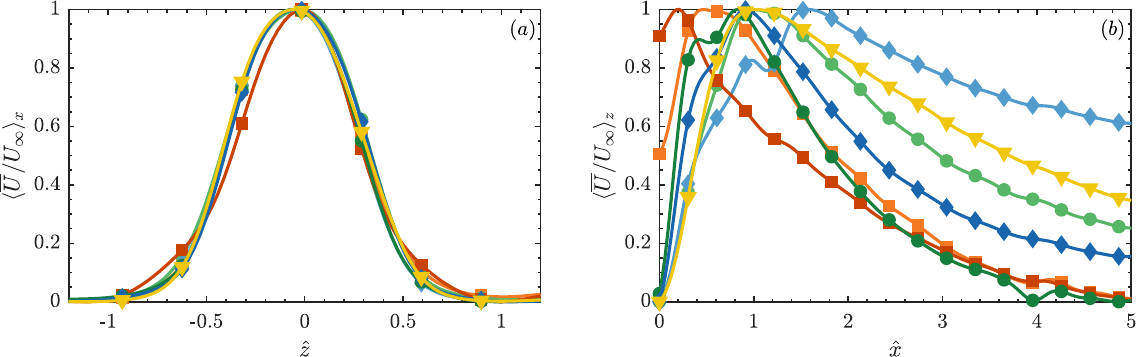}
    \caption{Streamwise velocity profiles at $y^+ = 220$ averaged along the (a) streamwise and (b) spanwise direction, and scaled with a min-max normalisation. Profiles are shown for both the steady-jet  actuation \sy{Yellow1}{t*} and all the selected pulsed-jet configurations for PIV measurements: $f = 225\mathrm{Hz}$ for \sy{f225_DC75}{d*}75\% DC, \sy{f225_DC50}{o*}50\% DC, \sy{f225_DC25}{s*}25\% DC; and $f = 150\mathrm{Hz}$ for \sy{f150_DC75}{d*}75\% DC, \sy{f150_DC50}{o*}50\% DC, \sy{f150_DC25}{s*}25\% DC.}
    \label{fig:PIV_profile_a}
\end{figure*}

The mean flow field features appear to be nearly independent of the actuation parameters. Thus, for the sake of clarity and conciseness, velocity results are reported as the streamwise and spanwise average profile to better envision the relevance of the actuation parameters. Similarly to the previous analysis on the Nusselt number (see figure~\ref{fig:similarity}), figures~\ref{fig:PIV_profile_a} depicts the streamwise velocity profiles averaged through the streamwise and spanwise directions for both the steady-jet actuation and the pulsed-jet configurations. The profiles are scaled with a \textit{min-max} normalisation to focus on the shape of the velocity distribution rather than on the magnitude. The scaled spanwise distribution in figure~\ref{fig:PIV_profile_a}(a) confirms the similarity of the actuation, with a shape that resembles that of the Nusselt number profiles in figure~\ref{fig:similarity}(a). This was an expected result since the spanwise distribution of the actuation is governed by the width of the slot jet, consequently affecting a very localised area downstream of the jet. On the contrary, the velocity distribution along the streamwise direction in figure~\ref{fig:PIV_profile_a}(b) highlights the considerable differences among the different control configurations. The peak of maximum velocity is displaced depending on the pulsation frequency and duty cycle, both showing a remarkable influence on the jet penetration.

\begin{figure*} 
    \centering
    \includegraphics[width=0.9\linewidth]{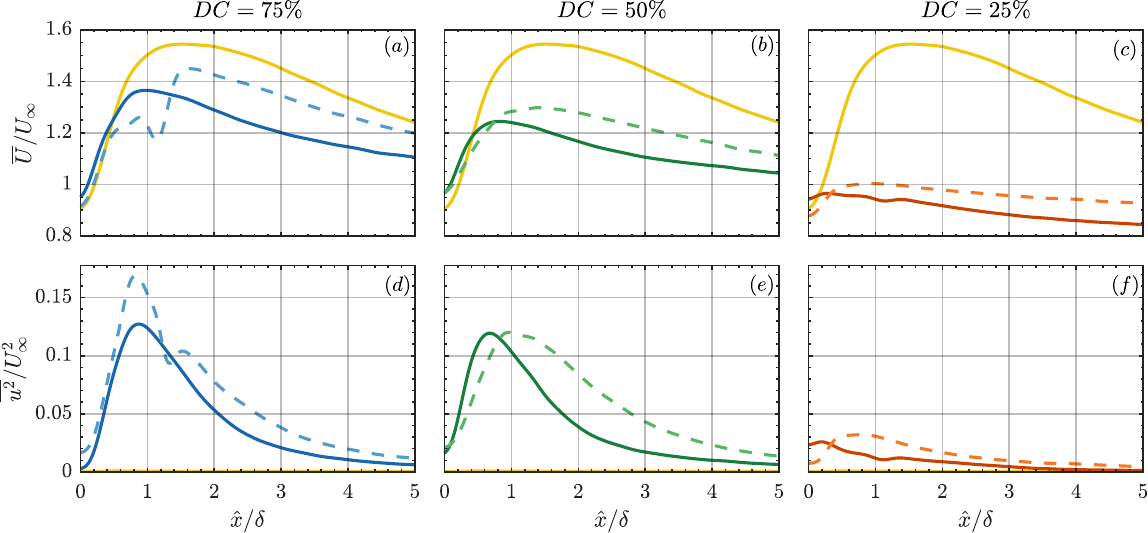}
    \caption{Streamwise velocity $\overline{U}$ (a-c) and fluctuations $\overline{u^2}$ (d-e) profiles at the mid-plane $z=0$ and $y^+ = 220$. Steady-jet \lcap{-}{Yellow1} actuation shown for reference. The line style differentiates between pulsation at $f = 225\mathrm{Hz}$ \lcap{-}{black} and $f = 150\mathrm{Hz}$ \lcap{--}{black}.}
    \label{fig:PIV_profile_b}
\end{figure*}

Figures~\ref{fig:PIV_profile_b}(a-c) and figures~\ref{fig:PIV_profile_b}(d-e) illustrate the streamwise distribution of the average streamwise velocity $\overline{U}$ and its fluctuations $\overline{u^2}$ for the different considered control parameters at the symmetry plane of the slot. The steady-jet actuation is included for reference, showing the maximum velocity peak at approximately $\hat{x} = 1.5\delta$ from the slot. Since the flow is not pulsed, the fluctuations are one order of magnitude lower than that of the pulsed jet. As expected, the duty cycle is the main promoter of jet penetration since it is a figure of merit of the mass and momentum injection into the flow. While all the cases show a velocity peak above the unperturbed case intensity, for $DC = 25\%$, the injected flow reaches the $y^+=220$ plane very weakly, rapidly adjusting to the unforced flow condition ($\overline{U}/U_\infty=0.8$), while for $DC = 75\%$ the flow is still considerably disturbed at the end of the measurement region. Apart from modifying the intensity of the velocity field, the duty cycle also alters the peak position, shifting it downstream for increasing the duty cycle and reaching its maximum displacement for the steady jet. This is possibly indicative of a different inclination of the jet as the duty cycle is increased. For the 6 different actuation configurations under evaluation, the position of the peak of maximum fluctuations coincides with the position of the peak of maximum streamwise velocity. The intensity of the peak of maximum fluctuations follows the same trend as the position of the peak of maximum streamwise velocity for different actuation parameters.  It is worth noting that the expected variance of the jet injection velocity based on the duty cycle becomes maximum for $DC = 50\%$, which explains the strong peak of $u^2$. The velocity peak position is also altered by the actuation frequency, slightly shifting upstream for increasing Strouhal number. Based on the definition of the effective duty cycle $DC^*$ in Eq.~\eqref{eq:effDC}, the injected mass flow slightly reduces with frequency, and hence the injection of mass into the mainstream.

A Proper Orthogonal Decomposition (POD) analysis is performed on the fluctuating velocity fields to shed more light on the flow structures describing the pulsed jet dynamics. The snapshot matrix $\boldsymbol{S}$ is built with the streamwise $u$ and spanwise $w$ fluctuating velocity components from the $n_t = 3000$ PIV snapshots, so that $\boldsymbol{s_i} = [\boldsymbol{s_i},\boldsymbol{s_i}]^T$. The Singular Value Decomposition (SVD) of the snapshot matrix reads
\begin{equation}
    \boldsymbol{X} = \sum_{n = 1}^{n_t} \sigma_n \boldsymbol{\psi_n}(t)\boldsymbol{\varphi_n}(x),
    \label{eq:SVD}
\end{equation}
being $\boldsymbol{\varphi_n}$, $\boldsymbol{\psi_n}$ and ${\sigma_n}$ the spatial function, temporal coefficients and singular value of the $n^{th}$ POD mode, respectively. The total number of modes is $n_t$ since the available number of grid points is larger than the number of snapshots.

\begin{figure}[t] 
    \centering
    \includegraphics[width=0.99\linewidth]{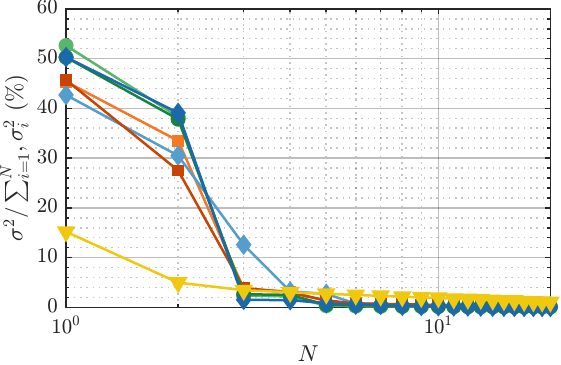}
    \caption{Spectrum of the POD modes of the streamwise $u$ and spanwise $w$ unsteady velocity component at $y^+ = 220$. Refer to figure \ref{fig:mdot} for marker and color legend.}
    \label{fig:POD_spectrum}
\end{figure}

The eigenspectrum is shown in figure~\ref{fig:POD_spectrum}. The square of the singular values $\sigma_i^2$ represent the contribution of each mode in building the in-plane turbulent kinetic energy. The eigenspectra are normalized with the respective sums of the squared singular values for each case, thus representing a fraction of energy collected in each mode. The pulsed JICF is dominated by the two first modes, which represent a considerable amount of the energy content. This phenomenon is common for periodic large-scale events. For the tested actuation cases, the first two modes are harmonically related and are in phase quadrature, similarly to a Fourier decomposition \citep{Raiola2016cylindertandem}. Modes 3 and 4 are higher harmonics of the former.

Taking advantage of this harmonic relation, the first two POD modes can be employed to derive a reduced-order model that provides phase information from non-time-resolved data. The following decomposition is considered \citep{ben2004reconstruction}:
\begin{equation}\label{eq:Ulor}
    U_{LOR} = \overline{U} + a_1(\theta)\boldsymbol{\varphi_1} + a_2(\theta)\boldsymbol{\varphi_2} + \sum_{n=3}^{N} a_n(\theta)\boldsymbol{\varphi_n}
\end{equation}
being $U_{LOR}$ the low-order reconstruction (LOR) of the velocity field, $\boldsymbol{\varphi_n}$ the $n^{th}$ spatial POD mode, and $a_n$ the $n^{th}$ temporal coefficient defined as,
\begin{equation}\label{eq:temp_modes}
    a_1(\theta) = \frac{\sqrt{2}}{\sqrt{N_t}}\sigma_1\sin{(\theta)}; ~~~ a_2(\theta) = \frac{\sqrt{2}}{\sqrt{N_t}}\sigma_2\cos{(\theta)}
\end{equation}
with $\theta$ being the period phase and $\sigma_n$ being the $n^{th}$ eigenvalue. Apart from the high energy content of the first two POD modes reported in figure~\ref{fig:POD_spectrum}, the coherent harmonic phenomena could be also verified by the Lissajous patterns illustrated when plotting the temporal coefficients scaled with the number of snapshots, namely $\boldsymbol{\psi}_1\sqrt{N_t/2}$ and $\boldsymbol{\psi}_2\sqrt{N_t/2}$. Although not shown here for simplicity, the coefficients distribute in the neighbourhood of a unit circle, confirming the validity of Eq.~\eqref{eq:Ulor}. This methodology has been previously used to describe the flow structure three-dimensional organisation in an injection system of an aero engine \citep{Ceglia2014}, the reconstruction of the wake behind tandem cylinders in a wall-bounded flow and a square rib with different incident angles \citep{Oudheusden2005phaseres}.

\begin{figure*}
    \centering
    \includegraphics[width=0.99\linewidth]{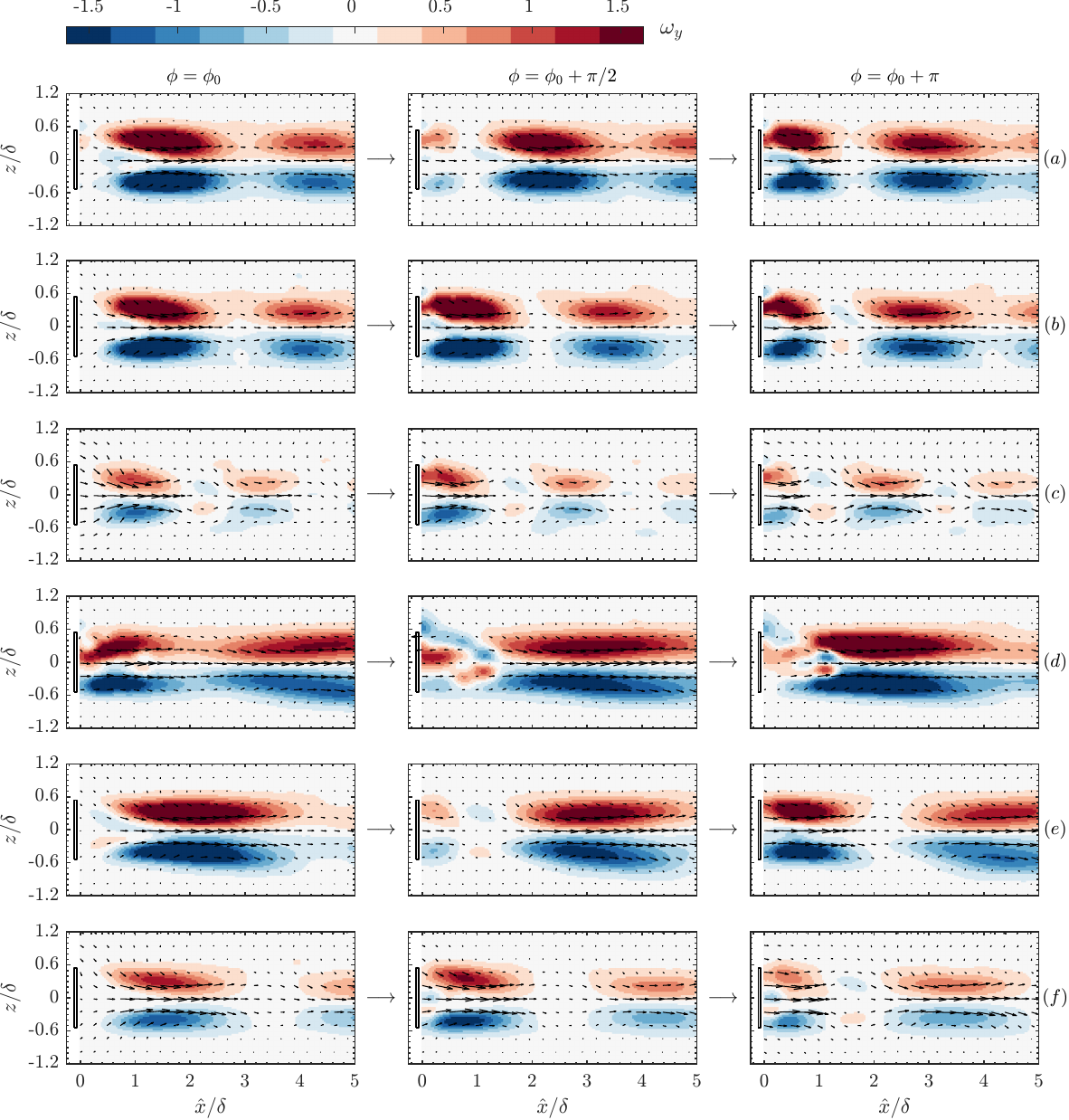}
    \caption{Low-Order Reconstruction of the wall-normal vorticity $\omega_y$ at $y^+ = 220$ for three different phases of the actuation cycle. The vector field depicts the velocity $U_{LOR}-U^*$ and $W_{LOR}$. Flow fields are shown for (a-c) the best frequency $f=225\mathrm{Hz}$ ($St = 0.97$) with $DC = 75,50,25\% $, and (d-f) $f=150\mathrm{Hz}$ ($St = 0.64$) with $DC = 75,50,25\% $. The location of the slot \lcap{-}{black} is included for reference. Note: the initial phase $\phi_0$ is arbitrary for each actuation case.}
    \label{fig:omega_lor_POD}
\end{figure*}

The streamwise and spanwise velocity components are reconstructed following Eq.~\eqref{eq:Ulor}, allowing to obtain the phase-resolved description of the jet-induced flow structures during one period of the actuation cycle. Figure~\ref{fig:omega_lor_POD} illustrates the wall-normal vorticity computed from $U_{LOR}$ and $W_{LOR}$ and normalised with $U_\infty$ and $\delta$. The superimposed vector field of $U_{LOR}-U^*$ and $W_{LOR}$ velocity is included on top. The vorticity maps are shown for three phases of the actuation cycle: initial phase $\phi_0$, followed by $\phi_0+\pi/2$ and $\phi_0+\pi$. It is to be noted that the initial phase $\phi_0$ is arbitrary since the PIV measurements are not synchronised with the jet pulsation. 

The wall-normal vorticity maps in figure~\ref{fig:omega_lor_POD} exhibit a common flow pattern independently of the control parameters. A pair of counter-rotating vortices show up at both sides of the slot, travelling downstream while promoting mixing. As it was discussed in figure~\ref{fig:PIV_ref} for the steady jet, the time-average flow field is characterised by a symmetric velocity distribution in which the vortices entrain surrounding air towards the centerline. The pulsation of the jet breaks the streamwise vortices, yielding the vorticity blobs depicted in figure~\ref{fig:omega_lor_POD}. Comparing the vorticity distribution for different duty cycles, it is observed that decreasing $DC$ translate into a reduction in the size of the induced structures, which appear to be more compact. On the other hand, there is a notable difference between the structures at the optimal actuation at $St = 0.97$ and the low-performance actuation at $St = 0.64$. The former describes more compact, smaller in size but stronger in intensity vorticity blobs travelling downstream, while the latter promotes the formation of a vortex that considerably elongates as it is convected downstream. Although the average injection velocity is the same independently of the actuation parameters, reducing the frequency (and augmenting the duty cycle) increases the pulse width, translating into a greater amount of injected air per pulse. The injected air mixes with the mainstream leading to streamwise rollers that are convected downstream while promoting the entrainment of air towards the mid-plane and towards the wall.

The LOR fields provide a direct mechanism to compute the size $L$, spacing $s$ and convective velocity $U_c$ of the induced structures. The results in table~\ref{tab:cohstruct} corroborate the aforementioned discussion based on the vorticity maps. The augmentation of the injection pulse width by altering either the frequency or the duty cycle is the main driver of the structure size, and consequently of the spacing. Regarding the convection velocity, the results show that the induced vortices travel faster than the mainstream. It is worth noting that, the streamwise velocity at the measurement plane for the unforced TBL is $U(y^+=220)=0.8U_\infty$. Thus, the jet action is not only promoting large-scale turbulent structures but also accelerates the bulk flow, which in combination yields a heat transfer enhancement at the wall.

\begin{table}[]
\caption{Characteristics of the large-scale structures induced by the jet actuation: size $L$, streamwise spacing $s$ and velocity of convection $U_c$.}\label{tab:cohstruct}
\begin{tabular}{@{}lcccccc@{}}
\toprule
$f$ {[}Hz{]} & \multicolumn{3}{c}{225} & \multicolumn{3}{c}{150} \\ \midrule
$DC$ (\%) & 75 & 50 & 25 & 75 & 50 & 25 \\ \midrule
$L/\delta$ & 1.36 & 1.34 & 1.15 & 2.34 & 2.22 & 1.84 \\
$s/\delta$ & 2.77 & 2.74 & 2.53 & 4.96 & 4.50 & 3.68 \\
$U_c/U_\infty$ & 1.18 & 1.17 & 1.08 & 1.41 & 1.28 & 1.05 \\ \bottomrule
\end{tabular}
\end{table}

There is a vast literature on jets in cross flow and the induced flow structures upon their modulation. Although refined during the last decades, there is consensus about the common flow organization induced by a round JICF. \citet{Sau2010optJICF} proposes a regime map depending on the stroke ratio and the ring velocity ratio. They conclude that for low ring velocity ratios, the jet tends to induce hairpin vortices. As the ring velocity ratio is increased, if the stroke ratio is small the jet induces a vortex ring tilted upstream, while higher stroke ratios yield a downstream tilted vortex ring accompanied by a trailing column of vorticity. These observations agree with \citet{Johari2006scaling}, in which the kind of flow structures are categorised based on the stroke ratio. According to this classification, the herein considered actuation would be characterised by vortex rings or puffs together with a trailing column. However, this study differs from the herein presented in the jet velocity ratio, the jet nozzle shape, and the inclination of the injection plane. The recent study by \citet{Steinfurth2021pulsedjet} copes with these aspects, deeply describing the flow topology and structures induced by an inclined, pulsed, slot jet in cross flow. They conclude that for an inclination angle of $30^\circ$, the flow structures appear to be independent of the velocity ratio. The jet flow attaches to the wall due to a local reduction in static pressure. The wall jet is characterised by two structures: a dominant large-scale leading vortex and a smaller trailing vortex tube. The leading vortex is the upper part of a half-ring vortex which is thinner compared to its legs extending in the streamwise direction. These streamwise vortices are the flow structures reported in figure~\ref{fig:omega_lor_POD}, which are responsible for the spanwise entrainment as well as for the formation and persistence of the leading vortex. For small velocity ratios, as the herein considered ($R = 0.7$), \citet{Steinfurth2021pulsedjet} report the cancellation of vorticity of the jet flow. However, wall-attached jets are always present. 

The production of vortex rings embedded in a turbulent boundary layer is not a common solution for the enhancement of momentum fluxes, being negligible if not detrimental. On the contrary, by inclining the injection plane, the jet fluid attaches to the wall while inducing a strong leading vortex that promotes streamwise momentum in the near-wall region. Following the description by \citet{Steinfurth2021pulsedjet}, three transport mechanisms are identified as major contributors: (1) low-momentum fluid is accelerated away from the wall by the wall jet and its leading vortex; (2) the raised fluid is replenished by the wall-attached jet, increasing the boundary-layer momentum; and, (3) high-momentum fluid from the free stream is entrained toward the wall in the trailing part of the wall jet fulfilling mass conservation. The above-discussed observations from the flow field measurements capture the legs of the half-ring vortex, which appear in the form of a streamwise counter-rotating vortex pair confining the wall jet in its central region.

A final comment is devoted to optimal pulsation. Among the different actuation parameters, there is a clear actuation frequency at which the Nusselt number is maximised as illustrated in figure~\ref{fig:Num} and figure~\ref{fig:eta}. The flow field measurement and the POD analyses highlight the difference in the flow topology induced by the jet actuation; however, there are no evident phenomena that explain the outstanding performance of $St \approx 1$. This actuation frequency seems to be a characteristic frequency of the dynamics of the system that, based on the definition of the Strouhal number, is related to the excitation frequency of the turbulent structures within the near-wall region. Among the wide variety of experimental investigations of pulsed jets in cross flow, there is no consensus on the optimal excitation frequency. \citet{kelso1996experimental} reports that the optimal jet pulsation should be close to the natural shedding frequency of the shear layer surrounding the jet. \citet{MCLOSKEY2002} explores the optimality among this natural frequency and its sub-harmonics; however, they conclude that the optimal frequency, if any, is apparatus dependent. On the other hand, there is a reasonable agreement on the fact that a reduced pulse width yields more compact vortex rings with higher penetration \citep[e.g.][]{kelso1996experimental,johari1999penetration, eroglu2001structure, MCLOSKEY2002, Johari2006scaling}. The identification of optimal pulsation frequency for mixing and/or heat transfer enhancement remains an open research question that requires further investigation, identifying the flow features, apparatus dependencies and scaling parameters defining the problem.

\section{Conclusions \label{s:Conclusions}}
The control of the convective heat transfer in a turbulent boundary layer by means of a pulsed, slot jet in crossflow has been investigated experimentally. A parametric study is carried out by varying the pulsation frequency and duty cycle. As expected, the injected mass of air by the jet is directly related to the duty cycle, hence being the main driver of the actuation cooling performance. On the contrary, the Nusselt number does not show a linear dependency with the actuation frequency. The actuation with the best absolute performance is $DC = 75\%$ and $f=225\mathrm{Hz}$ ($St \approx 1$); however, the inclusion of the volumetric flow rate of injected air to account for the efficiency of the system brings forward the control actions at the lowest duty cycle. It is concluded that, at a low duty cycle, a greater fraction of the pumping power is transformed into turbulent kinetic energy which translates into better utilisation of the injected volume of air within the TBL from a heat transfer enhancement perspective.

A simplified model is proposed for the estimation of decoupled effects of pulsation frequency and duty cycle on heat transfer enhancement. It is concluded that the effective duty cycle linearly decays with the pulsation frequency due to hardware limitations; however, the physical effect on heat transfer enhancement is directly linked to the duty cycle. It is worth remarking that this model is valid for the herein-considered conditions, mainly constrained by constant operating pressure of the pulsed jet. The assumed simplifications and mathematical formulation for other approaches, such as constant mass or volumetric flow rate, would require an adjustment on the considered control parameters.

The flow field measurements are performed for the frequency that maximises the Nusselt number ($St = 0.97 \approx 1$) and for another arbitrary case in which the performance of the heat transfer enhancement is considerably smaller ($St = 0.64 \approx 2/3$). The velocity fields together with the POD analysis conclude the generation of a pair of counter-rotating vortices convected downstream faster than the mainstream. Based on recent literature findings, it is concluded that these vortices at both sides of the slot jet are the legs of a half-ring vortex. The size and spacing of a large-scale structure depend on the actuation parameters while its convection velocity has been observed to be similar for all the tested cases. Conversely, the justification of the best pulsation frequency $St\approx 1$ remains an open question. Based on the adopted definition of the Strouhal number, the heat transfer rate is maximised at the frequency at which the large-scale turbulent structures within the boundary layer are convected in the near-wall region; nonetheless, further investigation is required to comprehend the flow mechanism and apparatus dependencies behind this outcome.

The results presented in this work are valid for the range of actuation parameters and the shape of the jet considered in this work. Yet, they highlight the features of a simple convective heat transfer enhancement strategy for turbulent flows. The simplified model for enhancement prediction and the identification of the involved flow structures in the heat transfer process pave the way toward the design of more complex and efficient control approaches, the derivation of analytical control laws and models, and the interpretability of the control performance.

\section*{Acknowledgments}
The work has been supported by the project ARTURO, ref. PID2019-109717RB-I00/AEI/10.13039/501100011033, funded by the Spanish State Research Agency. Paul Murphy is kindly acknowledged for his support during the experimental campaign.


\bibliographystyle{model1-num-names}
\bibliography{bib}

\end{document}